\documentclass[11pt,a4paper]{article}
\usepackage{jheppub}

\usepackage[utf8]{inputenc}
\usepackage{amsfonts,amssymb,amsmath,amsthm}
\usepackage{graphicx}
\usepackage{slashed}

\newcommand{\MM}{\mathcal{M}}
\newcommand{\ii}{\mathrm{i}}
\newcommand{\np}{\mathbf{n^+}}
\newcommand{\nm}{\mathbf{n^-}}
\newcommand{\nn}{\mathbf{n}}
\newcommand{\tr}{\mathrm{tr}}
\newcommand{\trg}{\mathrm{tr}_{\mathfrak{g}}}
\newcommand{\Ind}{\mathrm{Ind}(\slashed{D})}
\newcommand{\DD}{\mathcal{D}}
\newcommand{\IndD}{\mathrm{Ind}(\tilde\DD)}

\title{Index Theorems and Domain Walls}
\author[a,b]{Dmitri Vassilevich}
\affiliation[a]
{ CMCC, Universidade Federal do ABC,\\ Avenida dos Estados 5001, CEP 09210-580, Santo Andr\'e, SP, Brazil }
\affiliation[b]{Department of Physics, Tomsk State University,\\ 36 Lenin Ave, 634050 Tomsk, Russia}
\emailAdd{dvassil@gmail.com}

\abstract{The Atiyah-Patodi-Singer (APS) index theorem relates the index of a Dirac operator to an integral of the Pontryagin density in the bulk (which is equal to global chiral anomaly) and an $\eta$ invariant on the boundary (which defines the parity anomaly).  We show that the APS index theorem holds for configurations with domain walls that are defined as surfaces where background gauge fields have discontinuities.}

\begin{document}

\maketitle

\section{Introduction}\label{sec:intro}
In a series of papers \cite{Atiyah:1975jf,Atiyah:1976jg,Atiyah:1980jh} Atiyah, Patodi and Singer (APS) demonstrated the following formula for the index of a Dirac operator $\slashed{D}$  on an even-dimensional Riemannian manifold $\MM$ with boundary:
\begin{equation}
\Ind = \int_\MM P(x)d^nx  - \tfrac 12 \eta (0,\DD)\,, \label{APS}
\end{equation}
$n=\mathrm{dim}\, \MM$.
Here $P(x)$ is a Pontryagin type density, and $\eta (0,\DD)$ is the spectral asymmetry of some auxiliary Dirac operator defined on the boundary $\partial\MM$. The Dirac operator is subject to non-local APS boundary conditions: the modes corresponding to non-negative eigenvalues of $\DD$ have to vanish on $\partial\MM$. We postpone precise definitions to Sections \ref{sec:in} and 
\ref{sec:APS}. A detailed introduction to index theorems may be found in \cite{Gilkey:1984} and a physicist-friendly survey - in \cite{Witten:2015aba}.

The APS formula (\ref{APS}) is quite remarkable from physical point of view. It relates the index to the chiral anomaly in the bulk (given by the Pontryagin term) and to the parity anomaly \cite{Niemi:1983rq,Redlich:1983dv} in a boundary theory (given by the $\eta$-function \cite{AlvarezGaume:1984nf}). The APS formulas were used in physics already in 1980 \cite{Hortacsu:1980kv}. Some implications of this and similar relations to topological materials have been discussed in \cite{Witten:2015aba} and \cite{Chu:2018ksb}. Some other aspects of boundary contributions to the anomalies were considered relatively recently in \cite{Herzog:2015ioa,Fursaev:2015wpa,Solodukhin:2015eca,Kurkov:2017cdz,Jensen:2017eof,Kurkov:2018pjw} as also in much earlier works, see \cite{Moss:1994jj,Esposito:1997wt,Kirsten:2001wz,Vassilevich:2003xt,Marachevsky:2003zb} and references therein. 

Unfortunately, any direct use of the APS formula in physics is limited by the fact that it needs non-local boundary conditions on $\partial\MM$. One cannot define the index with local boundary conditions since the latter mix the modes of different chiralities \cite{Marachevsky:2003zb,Witten:2015aba,Jensen:2017eof}. Thus, if one likes to have both the index theorem and the locality, one has to look for a different setup.

In this paper, we consider 4D fermions on a background with a domain wall which is understood as a 3D submanifold where background gauge fields have a discontinuity (quite in the spirit of domain walls in ferromagnets). The fermions are supposed to be continuous across the wall. Such backgrounds admit a split of the spectrum of $\slashed{D}$ into modes of different chiralities and thus allow to compute $\Ind$. Our main results is that a close analog of the APS formula, see Eq.\ (\ref{APSwall}) below, holds for this case\footnote{A different proposal for an extension of APS theorem to domain wall was made in \cite{Fukaya:2017tsq}. That paper treated a massive Dirac operator that did not admit eigenmodes of definite chirality. The index was defined as a phase of some functional determinant. The approach of \cite{Fukaya:2017tsq} has little in common with the one adopted here.}. To prove this, we compute both sides of (\ref{APSwall}) by using the heat kernel methods. 

This paper is organized as follows. The next section contains main definitions and describes geometry of the problem. In Section \ref{sec:in}, we compute the index on domain wall background. In Section \ref{sec:APS}, the APS index theorem for domain walls is derived. Some explicit examples are considered in Section \ref{sec:ex}. Section \ref{sec:dis} contains concluding remarks.

\section{The setup}\label{sec:set}
Let $\MM$ be a smooth compact four-dimensional manifold, and let $\Sigma$ be a smooth codimension one submanifold that separates $\MM$ in two parts, $\MM_+$ amd $\MM_-$. In the Atiayh-Singer approach, the index is associated with some elliptic complex over $\MM$, which is the spin complex in our case. It consists of a spin-bundle $\mathcal{S}$ with two subbundles $\mathcal{S}_R$ and $\mathcal{S}_L$ of chiral spinors, $\mathcal{S}=\mathcal{S}_R \oplus \mathcal{S}_L$, and of a Dirac operator $\slashed{D}$. The operator $\slashed{D}$ has to map sections of $\mathcal{S}_R$ to sections of $\mathcal{S}_L$, and vice versa. This structure imposes many restrictions on the operator $\slashed{D}$ and on the form of matching conditions on $\Sigma$. Since $\slashed{D}$ has to anti-commute with the chirality matrix $\gamma^5$, we are left with interaction with an external metric, a gauge field, and an axial vector field. To avoid unnecessary complications, we suppose that the metric $g_{\mu\nu}$ is flat, and there is no axial vector field. Thus, 
\begin{equation}
\slashed{D}=\ii \gamma^\mu \nabla_\mu\,,\qquad \nabla_\mu = \partial_\mu +A_\mu \,.\label{Dnab}
\end{equation}
Here $A_\mu$ is a non-abelian gauge field with the field strength $F_{\mu\nu}=\partial_\mu A_\nu + \partial_\nu A_\mu + [A_\mu,A_\nu]$.
The gamma matrices satisfy $\gamma^\mu\gamma^\nu + \gamma^\nu\gamma^\mu =2g^{\mu\nu}$ and
\begin{equation}
\tr\, \bigl( \gamma^5\gamma^\nu\gamma^\nu\gamma^\rho\gamma^\sigma \bigr)=4\epsilon^{\mu\nu\rho\sigma} \label{LC}
\end{equation}
with the Levi-Civita tensor $\epsilon^{\mu\nu\rho\sigma}$.

We suppose that the metric $g$ on $\MM$ is smooth across $\Sigma$, and that extrinsic curvature of $\Sigma$ vanishes (meaning that $\Sigma$ is a totally geodesic submanifold in $\MM$). The induced metric on $\Sigma$ will be denoted by $h$.
We denote by $x^a$, $a=1,2,3$, the coordinates on $\Sigma$ and by $\np$ and $\nm$ the unit normal vectors to $\Sigma$ that are pointing inside $\MM_+$ and $\MM_-$, respectively. Since the geometry is smooth across $\Sigma$, we have $\np=-\nm \equiv \nn$. Superscripts $+$ and $-$ will be used to denote the limiting values of various fields on $\Sigma$ from $\MM_+$ and $\MM_-$, respectively.  

We allow the tangential components of gauge field to have a discontinuity on $\Sigma$
\begin{equation}
B_a=A_{a}^+-A_{a}^-\,. \label{Ba}
\end{equation} 
We assume that gauge connections induced on $\Sigma$ from $\MM_+$ and $\MM_-$ are compatible, i.e. they are connections in the same vector bundle. This amounts to saying that $B_a$ is a vector (since it is a difference of two connections). We also allow normal components $A_\nn$ of gauge connection to jump on $\Sigma$, though this is by far less important since $A_\nn^+$ and $A_\nn^-$ can be removed by a continuous gauge transformation. We shall see below, that these normal components do not enter the expression for $\Ind$, see Eqs.\ (\ref{a4a}) - (\ref{a4c}).

One has to define some matching conditions for the spinors $\psi$ on the interface $\Sigma$. We request the continuity,
\begin{equation}
\psi^+=\psi^- \label{match1}
\end{equation}
Since the operator $\slashed{D}$ has to be hermitian, $\slashed{D}\psi$ should also be continuous across $\Sigma$. This gives us a matching condition for the normal derivatives
\begin{equation}
\nabla_{\np}\psi^+ + \nabla_{\nm}\psi^- =U\psi, \qquad U=-\gamma^\nn \gamma^a B_a \label{match2}
\end{equation}

Possible generalizations of this setup will be considered in Section \ref{sec:dis}.

\section{Computation of the Index}\label{sec:in}
The index of a Dirac operator is defined as a difference between the numbers of zero modes having positive and negative chiralities. It cam be written as a functional trace
\begin{equation}
\Ind =\mathrm{Tr}\, \left( \gamma^5 e^{-t\slashed{D}^2} \right), \qquad t>0.\label{Ind1}
\end{equation}
This expression can be evaluated with the help of the heat kernel expansion. For any operator of Laplace type 
\begin{equation}
L=-(\nabla_\mu \nabla^\mu +E) \label{L}
\end{equation}
 and any zeroth order operator $Q$ there is an asymptotic expansion 
\begin{equation}
\mathrm{Tr}\, \left( Q e^{-tL} \right) \simeq \sum_{k=0}^\infty t^{\frac {n-4}2} a_k(Q,L), \qquad t\to +0. \label{asymp}
\end{equation}
Since the left hand side of (\ref{Ind1}) does not depend on $t$, only the constant term in (\ref{asymp}) contributes to the index,
\begin{equation}
\Ind = a_4(\gamma^5, \slashed{D}^2)\,. \label{Ind2}
\end{equation}

The heat kernel coefficients for matching conditions of the type (\ref{match1}), (\ref{match2}) were computed with increasing degree of generality in the papers \cite{Bordag:1999ed,Moss:2000gv,Gilkey:2001mj}. These coefficients are integrals over $\MM$ and $\Sigma$ of local invariant polynomials constructed from the curvatures, the filed strength, $E$, $U$, $B_a$ and their derivatives. In our case,
\begin{equation}
E=\tfrac 14 [\gamma^\mu,\gamma^\nu] F_{\mu\nu}\,, \label{E}
\end{equation}
while other relevant quantities are given in (\ref{Ba}) and (\ref{match2}). The computation of Ref.\ \cite{Gilkey:2001mj} was done for $Q$ being a function without any matrix structure. Thus, in principle, the expressions containing commutators of $Q$ with other quantities may appear in the answer with yet undetermined coefficients. Fortunately, for $Q=\gamma^5$ all such commutators vanish, and we can use the results of \cite{Gilkey:2001mj}. The relevant terms in $a_4$ read
\begin{eqnarray}
&&a_4(\gamma^5,\slashed{D}^2)= \frac 1{12(4\pi)^2} \left[ \int_{\MM} \sqrt{g} d^4x\, \tr\, \gamma^5\bigl( 2\nabla^\mu\nabla_\mu E +6E^2 +F_{\mu\nu}F^{\mu\nu} \bigr) \right. \nonumber\\
&& \qquad +  \int_\Sigma \sqrt{h} d^3x \, \tr\, \gamma^5 \bigl( 2(\nabla_{\np}E^+ +\nabla_{\nm}E^-)-2B^a(F^+_{a\np}-F^-_{a\nm}) \nonumber\\
&& \qquad \left.  -2U^3-6U(E^++E^-) -2UB_aB^a -2\nabla^a\nabla_a U \bigr) \right]\,. \label{a4}
\end{eqnarray}
Here the trace is taken over all indices (gauge and spinor ones).

After an integration by parts, the term $2\nabla^\mu\nabla_\mu E$ in the bulk cancels the terms with normal derivatives of $E$ on $\Sigma$. The terms with $F_{\mu\nu}F^{\mu\nu}$, $B^a(F^+_{a\np}-F^-_{a\nm})$ and $UB_aB^a$ vanish after taking the trace over spinor indices. The term $\nabla^a\nabla_a U$ does not depend on the choice of the boundary covariant derivative, which may be either $\nabla^+_a$ or $\nabla^-_a$, and also vanishes after taking the trace. Remaining terms can be easily computed.
\begin{eqnarray}
&&\tr\, \gamma^5 E^2= \epsilon^{\mu\nu\rho\sigma}\trg F_{\mu\nu}F_{\rho\sigma}\,,\nonumber\\
&&\tr\, \gamma^5 U^3= 4\epsilon^{\nn abc} \trg B_aB_bB_c\,,\nonumber \\
&&\tr\, \gamma^5 U (E^++E^-) = -2\epsilon^{\nn abc} \trg B_a (F_{bc}^+ + F_{bc}^-)\,,\label{traces}
\end{eqnarray}
where $\trg$ means a trace over gauge indices only.
\begin{subequations}
\begin{align}
&\Ind = a_4(\gamma^5,\slashed{D}^2)= \frac 1{32\pi^2} \int_{\MM} \sqrt{g} d^4x\, \epsilon^{\mu\nu\rho\sigma} \trg\,  F_{\mu\nu}F_{\rho\sigma} \label{a4a} \\
&\quad +\frac 1{8\pi^2} \int_\Sigma \sqrt{h} d^3x \, \epsilon^{\nn abc}\trg\, \bigl( A_a^+\partial_b A_c^++\tfrac 23
A_a^+A_b^+A_c^+ - A_a^-\partial_b A_c^- - \tfrac 23 A_a^-A_b^-A_c^- \bigr) \label{a4b}\\
&\quad +\frac 1{8\pi^2} \int_\Sigma \sqrt{h} d^3x \, \epsilon^{\nn abc}\trg\, \bigl( A_a^+\partial_b A_c^- -
A_a^-\partial_b A_c^+ \bigr).\label{a4c}
\end{align}
\end{subequations}
As expected (cf Sec.\ \ref{sec:set}) this expression does not contain normal components of the gauge field on $\Sigma$. The expression on the first line above, (\ref{a4a}) is an integral of the Pontryagin density,
\begin{equation}
\int_\MM P(x)d^4x = \frac 1{32\pi^2} \int_{\MM} \sqrt{g} d^4x\, \epsilon^{\mu\nu\rho\sigma} \trg\,  F_{\mu\nu}F_{\rho\sigma} \,,\label{Pont}
\end{equation} 
which gives the instanton number if there are neither boundaries nor domain walls. The second line, (\ref{a4b}) is a combination of two Euclidean Chern-Simons actions
\begin{equation}
\frac \ii{2\pi} \bigl[ \mathrm{CS}_+(A^+) + \mathrm{CS}_-(A^-) \bigr] \,.\label{CS}
\end{equation}
The Chern-Simons action depends on the orientation of underlying manifold. The subscripts $\pm$ in the formula above stress that $\mathrm{CS}_+$ and $\mathrm{CS}_-$ are computed with the orientation of $\Sigma$ induced from $\MM_+$ and $\MM_-$, respectively.
\begin{equation}
\mathrm{CS}_\pm(A)=\pm \mathrm{CS}(A), \qquad \mathrm{CS}(A)\equiv \frac \ii{4\pi} \int_\Sigma \sqrt{h} d^3x \, \epsilon^{abc}\trg\, \bigl( A_a\partial_b A_c+\tfrac 23 A_aA_bA_c \bigr)\label{CSpm}
\end{equation}
and
\begin{equation}
\epsilon^{abc}=-\epsilon^{\np abc}=\epsilon^{\nm abc}. \label{epep}
\end{equation}
We included an imaginary unit in the definition of Chern-Simons action to ensure that it gives a phase to the Euclidean partitions function $Z\propto \exp (-\kappa \mathrm{CS}(A))$ for a real Chern-Simons level $\kappa$.

Note, that though one is allowed to integrate by parts in the expressions like $\trg (\nabla_\mu C^\mu)$ if $C$ is a vector, one cannot do this in general if $C$ is a gauge connection. The reason for this difference is in different transition  rules from one coordinate chart to another for vector-like quantities and for the gauge fields. There are, however, exceptions. One is related to small localized  variations of the connection $A_\mu \to A_\mu + \delta A_\mu$. If the support of $\delta A_\mu$ is small enough to fit into a single coordinate chart, one can freely integrate by parts in the expressions that are linear in $\delta A_\mu$. Thus, linear local variations vanish of the volume term (\ref{a4a}) taken together with the Chern-Simons surface term (\ref{a4b}), and separately - of the third line (\ref{a4c}).

If the line bundle is topologically trivial, the gauge connection $A_\mu$ may be defined globally on $\MM_+$ and $\MM_-$. Thus, the integrations by parts are allowed in (\ref{a4a}) - (\ref{a4c}), and $\Ind$ vanishes, as expected. If the induced bundle on $\Sigma$ has a trivial topology, the expression (\ref{a4c}) vanishes.  

Note, that \emph{topological invariants} have to be invariant under the transformations $A_\mu \to A_\mu +V_\mu$, where $V_\mu$ is a gauge and diffeomorphism vector, not necessarily localized. The topological invariance means considerably more that just the invariance under local variations. The index is a topological invariant, while the individual parts of the expression (\ref{a4a}) - (\ref{a4c}) are not.

\section{The APS Index Theorem} \label{sec:APS}
For the discussion in this section it is convenient to fix a particular representation of the $\gamma$-matrices:
\begin{equation}
\gamma^a=\left( \begin{array}{cc} 0 & 1 \\ 1 & 0 \end{array} \right) \otimes \sigma^a,\qquad
\gamma^\nn=\gamma^4=\left( \begin{array}{cc} 0 & i \\ -i & 0 \end{array} \right) \otimes \mathrm{id},\qquad
\gamma^5=\left( \begin{array}{cc} 1 & 0 \\ 0 & -1 \end{array} \right) \otimes \mathrm{id},
\end{equation}
where $\sigma^a$'s are the Pauli matrices, and $\mathrm{id}$ is a $2\times 2$ identity matrix. The Dirac operator takes the following form
\begin{equation}
\slashed{D}=\left( \begin{array}{cc} 0 & -\nabla_\nn + \DD \\
\nabla_\nn +\DD & 0 \end{array} \right), \qquad \DD=i\sigma^a \nabla_a \,.\label{DD}
\end{equation}
The operator $\DD$ is the one that appears in the APS index formula (\ref{APS}). It can also be defined in a way that does not depend on representations of the $\gamma$-matrices \cite{Gilkey:1984}. In our setting, there are two sides of $\Sigma$ and two sets of the gauge fields. Thus we define two Dirac operators on the brane,
\begin{equation}
\DD^+:=\DD (A^+)\,, \qquad \DD^-:=-\DD (A^-)\,. \label{DDDD}
\end{equation}
The minus sign in front of $\DD^-$ appears since $\nn=\np=-\nm$.  

The $\eta$ function of $\DD$ is defined as a sum over eigenvalues $\lambda$,
\begin{equation}
\eta (s,\DD) = \sum_{\lambda >0} \lambda^{-s} - \sum_{\lambda <0} (-\lambda)^{-s} .\label{etas}
\end{equation}
These sums are absolutely convergent for $\Re s$ sufficiently large. From this region, $\eta(s,\DD)$ can be continued to the whole complex plane as a meromorphic function. Let us stress, that zero eigenvalues are not included in (\ref{etas}), see \cite{Gilkey:1984}. One of the features of $\eta(0,\DD)$ is that it jumps by $\pm 2$ when an eigenvalue of $\DD$ crosses the origin. Therefore, in many cases in physics \cite{Witten:2015aba} and mathematics \cite{Gilkey:1984} the value $\eta(0,\DD)$ is reduced modulo $2\mathbb{Z}$. Since the integrals in (\ref{a4b}) and (\ref{a4c}) are smooth in $A$, we need a function $\tilde\eta(\DD)$ that coincides with $\eta(0,\DD)$ modulo $2\mathbb{Z}$,
\begin{equation}
\tilde\eta (\DD)= \eta(0,\DD) \quad \mbox{mod}\ 2\mathbb{Z} \,,\label{tileta}
\end{equation}
but depends smoothly on $A$. 

Let us now evaluate $\eta(0,\DD)$. On a general background, just a variation of $\eta(0,\DD)$ with respect to a variation $A_\mu \to A_\mu + \delta A_\mu$ can be computed. This is a rather standard computation in spectral geometry, see \cite{Gilkey:1984}. We refer to \cite{Fursaev:2011zz} for details.
\begin{equation}
\delta \eta (0,\DD)=\frac 1{4\pi^2} \int_\Sigma d^3x \sqrt{h} \epsilon^{abc} \trg \bigl( (\delta A_a)\ F_{bc} \bigr) \label{vareta}
\end{equation}
where we used $\tr (\sigma^a \sigma^b \sigma^c)=2\ii \epsilon^{abc}$.
We stress, that the variation $\delta A_\mu$ in Eq.\ (\ref{vareta}) does not need to be local, but has to be a vector. Eq.\ (\ref{vareta}) is valid for the variations $\delta A$ under which the eigenvalues of $\DD$ do not cross the origin. However, since $\eta(0,\DD)$ is continuous modulo $2\mathbb{Z}$, one can write the same formula as (\ref{vareta}) for $\tilde\eta$,
\begin{equation}
\delta \tilde\eta (\DD)=\frac 1{4\pi^2} \int_\Sigma d^3x \sqrt{h} \epsilon^{abc} \trg \bigl( (\delta A_a)\ F_{bc} \bigr) \label{vartileta}
\end{equation}
which is now valid for arbitrary variations to the linear order, cf.\ \cite{Gilkey:1984}. 

Let us consider the second variation of $\tilde\eta$ under two consecutive variations of $A$,
\begin{equation}
\delta^2 \tilde\eta (A)= \frac 1{2\pi^2} \int_\Sigma d^3x \sqrt{h} \epsilon^{abc} \trg \bigl( (\delta A_a)_1 \nabla_b (\delta A_b))_2 \bigr)\,.\label{del2} 
\end{equation}
If both variations, $(\delta A)_1$ and $(\delta A)_2$ are gauge and diffeomorphism vectors, one can integrate by parts in (\ref{del2}) to show that the second variation is symmetric. Thus, the variation equation (\ref{vartileta}) is integrable. We like to use this equation to evaluate $\tilde\eta (\DD^+)+\tilde\eta(\DD^-)=\tilde\eta (\DD(A^+))-\tilde\eta(\DD(A^-))$. Let us define a homotopy $A(t)_a=A^-_a +tB^a$, $t\in [0,1]$, so that $A(0)_a=A^-_a$ and $A(1)_a=A^+_a$. Then,
\begin{equation}
F(t)_{bc}=F_{bc}^-+t(\nabla^-_bB_c-\nabla^-_cB_b)+t^2[B_b,B_c]
\end{equation}
and, by (\ref{vartileta}),
\begin{equation}
\partial_t \tilde\eta(\DD(t))=\frac 1{4\pi^2} \int_\Sigma d^3x \sqrt{h} \epsilon^{abc} \trg \bigl( B_a\ F(t)_{bc} \bigr)\,. \label{dereta}
\end{equation}
After integrating (\ref{dereta}) over $t$ between $0$ and $1$ (and some elementary algebra), we obtain
\begin{eqnarray}
&&\tilde\eta(\DD^+)+\tilde\eta(\DD^-)=\frac 1{4\pi^2} \int_\Sigma \sqrt{h} d^3x \, \epsilon^{abc} \trg\, \bigl( A_a^+\partial_b A_c^++\tfrac 23
A_a^+A_b^+A_c^+ \nonumber\\ 
&&\qquad\qquad - A_a^-\partial_b A_c^- - \tfrac 23 A_a^-A_b^-A_c^- 
+ A_a^+\partial_b A_c^- -
A_a^-\partial_b A_c^+ \bigr).\label{tileta2}
\end{eqnarray} 
Then, by comparing this equation to (\ref{a4a}) - (\ref{a4c}) and taking into account (\ref{epep}), one gets 
\begin{equation}
\Ind =\int_\MM P(x)d^4x  - \tfrac 12 \bigl( \tilde\eta (\DD^+)+\tilde\eta(\DD^-) \bigr). \label{APSwall}
\end{equation} 
This is the APS index theorem for domain walls and the main result of this paper.

\section{Examples}\label{sec:ex}
To see better how the things work, let us consider two examples when $\tilde\eta(\DD^+)$ and $\tilde\eta(\DD^-)$ can be computed separately.

Our first example is a \emph{topologically trivial} gauge field on $\Sigma$. Then $A$ can be connected by a smooth homotopy to $A=0$.
The variation (\ref{vartileta}) is easily integrated to give 
\begin{equation}
\tilde\eta(\DD)= -\frac \ii{\pi} \mathrm{CS}(A) + \mathcal{C}\,, \label{etatriv}
\end{equation}
where $\mathcal{C}$ is an $A$-independent constant which is defined by the geometry of $\Sigma$. This constant is canceled in the combination $\tilde\eta (\DD^+)+\tilde\eta(\DD^-)$. As we mentioned at the end of Section \ref{sec:in}, the expression (\ref{a4c}) vanishes for topologically trivial configurations, and thus (\ref{APSwall}) is confirmed. 

As a second example, let us consider a \emph{topologically non-trivial} configuration with a $U(1)$ gauge group, $A_\mu=\ii\mathcal{A}_\mu$. This example is inspired by \cite{Deser:1997gp}, shares some  similarities with \cite{Fosco:2017vxl} and is very close to the one considered in \cite{Kurkov:2017cdz} (see also \cite{Ma:2018efs}). Let us take $\Sigma = \tilde\Sigma\times S^1$. The surface element on $\tilde\Sigma$ will be denoted by $\sqrt{\tilde h}$. Let the gauge field along unit circle $S^1$ be a constant, $\mathcal{A}_3=a$, and assume that $\mathcal{A}_1$ and $\mathcal{A}_2$ do not depend on $x^3$. Then,
\begin{equation}
\DD = \ii \sigma^3 (\partial_3 +\ii a) +\tilde\DD \,.\label{DDtDD}
\end{equation}
$\tilde\DD=\ii \sigma^\alpha \nabla_\alpha$, $\alpha=1,2$,  is the standard Dirac operator on $\tilde\Sigma$. Let us consider its' spectrum $\tilde\DD=\mu \psi(\mu)$. The operator $\tilde\DD$ anti-commutes with $\sigma^3$. Thus, all non-zero eigenmodes appear in pairs with opposite eigenvalues $\pm\mu$. One can show \cite{Kurkov:2017cdz} that the corresponding part of spectrum of $\DD$ is also symmetric. Consequently, only the eigenmodes of $\tilde\DD$ that have $\mu=0$ may contribute to the spectral asymmetry $\eta(0,\DD)$. Zero modes of $\tilde\DD$ are not paired. Let $N_+$ (resp., $N_-$) be the number of zero modes of $\tilde\DD$ corresponding to the eigenvalue $+1$ (resp., $-1$) of $\sigma^3$. Then
\begin{equation}
\eta (s,\DD)=(N_+-N_-) \eta (s, \ii (\partial_3 +\ii a)) \,.\label{eta3}
\end{equation}
This formula is a particular case of a more general result for $\eta$ functions on product manifolds \cite[Lemma 4.3.6.]{Gilkey:1984}.
The $\eta$ function on the right hand side of this equation has been computed in \cite{Gilkey:1984,Deser:1997gp,Kurkov:2017cdz}. For periodic boundary conditions on $S^1$ it reads
\begin{equation}
\eta (s, \ii (\partial_3 +\ii a))=\zeta_R(s,1-\bar a)-\zeta_R(s,\bar a)\,,\label{eta4}
\end{equation}
where $\bar a=a-\left\lfloor a \right\rfloor$ is the non-integer part of $a$, and $\zeta_R$ is the generalized Riemann (Hurwitz) $\zeta$ function. At $s=0$
\begin{equation}
\eta (0, \ii (\partial_3 +\ii a)) =2\bar a -1 \,.\label{eta5}
\end{equation}
To get a formula for \emph{anti-periodic} conditions on $S^1$, one has to shift $\bar a$ by $1/2$, so that
\begin{equation}
\eta (0, \ii (\partial_3 +\ii a)) =2\bar a  \,.\label{eta6}
\end{equation}

The expression $N_+-N_-$ is the index of $\tilde\DD$, which is defined by the well known formula through the Pontryagin number
\begin{equation}
\IndD = a_2(\sigma^3,\tilde\DD^2)=-\frac 1{4\pi} \int_{\tilde \Sigma} d^2x \sqrt{\tilde h} \epsilon^{\alpha\beta}
\mathcal{F}_{\alpha\beta} \label{Pont1}
\end{equation}
(see, e.g., \cite{Vassilevich:2003xt}). Here $\mathcal{F}_{\alpha\beta}=\partial_\alpha \mathcal{A}_\beta -\partial_\beta \mathcal{A}_\alpha$ and $\epsilon^{\alpha\beta}=\epsilon^{\alpha\beta 3}$. Thus, collecting everything together we obtain
\begin{eqnarray}
&& \eta(0,\DD)=-\frac{2\bar a -1}{4\pi} \int_{\tilde \Sigma} d^2x \sqrt{\tilde h} \epsilon^{\alpha\beta}
\mathcal{F}_{\alpha\beta} \quad \mbox{(periodic)}\,,\label{etaper}\\
&& \eta(0,\DD)=-\frac{2\bar a}{4\pi} \int_{\tilde \Sigma} d^2x \sqrt{\tilde h} \epsilon^{\alpha\beta}
\mathcal{F}_{\alpha\beta} \quad \mbox{(anti-periodic)}\label{etaanti}
\end{eqnarray}
for periodic\footnote{Note, that since $\IndD$ may in principle be any integer, $\eta(0,\DD)$may be arbitrarily large. This shows that the formula for $\eta$ guessed in \cite{Fukaya:2017tsq} basing on a 1-dimensional example is incorrect in higher dimensions.} and anti-periodic conditions, respectively. The smooth extension of $\eta$ is achieved by replacing $\bar a$ by $a$ in the formulas above,
\begin{equation}
\tilde\eta (\DD)=\eta(0,\DD)\vert_{\bar a \to a}\,.\label{teta}
\end{equation} 

One can easily check that for anti-periodic conditions, for example,
\begin{equation}
\tilde\eta (\DD) = -\frac{2\ii}{\pi} \mathrm{CS}[i\mathcal{A}] \,.\label{2CS}
\end{equation}
The coefficient in front of $ \mathrm{CS}[A]$ in (\ref{2CS}) is twice the one in (\ref{etatriv})\footnote{Non-uniqueness of the coefficient in front of CS action in the parity anomaly was discussed in the context of lattice models in \cite{Coste:1989wf}.}, though $\eta(0,\DD)$ and $\tilde\eta(\DD)$ still satisfy the same variation equations (\ref{vareta}) and (\ref{vartileta}) as the $\eta$-function for topologically trivial backgrounds, Eq.\ (\ref{etatriv}). The reason for discrepancy is that the well-known formula
\begin{equation}
\delta \mathrm{CS}[i\mathcal{A}]= - \frac \ii{2\pi} \int \sqrt{h} \epsilon^{abc}(\delta\mathcal{A}_a)\partial_b\mathcal{A}_c \label{varCS}
\end{equation}
is valid on general backgrounds for localized variations only (cf. discussion at the end of Section \ref{sec:in}). Thus, it is \emph{not} true that $\eta(0,\DD)$ is always given by the Chern-Simons term (\ref{etatriv}) with the same universal coefficient plus a topological invariant\footnote{Let us stress that we always use the definition (\ref{CSpm} for the Chern-Simons action. One can use another definition through the Pontryagin integral, see Eq.\ (2.49) in \cite{Witten:2015aba}. This latter action differs from ours on topologically non-trivial backgrounds and has somewhat different properties.}. The case considered here can serve as a counterexample: the difference between (\ref{etaper}) or (\ref{etaanti}) and (\ref{etatriv}) is not a topological invariant since there is a smooth homotopy (continuous variations of $a$) that changes this difference. However, it is true that
\begin{equation}
\tilde \eta (\DD) = -\frac{\ii}{\pi} \mathrm{CS}[i\mathcal{A}] +\mathcal{L}(\mathcal{A})\,,
\label{etaL}
\end{equation}
where $\mathcal{L}(\mathcal{A})$ is a functional having vanishing local variations with respect to $\mathcal{A}$. For the example considered here, 
\begin{equation}
\mathcal{L}(\mathcal{A})=-\frac \ii{4\pi} \int_\Sigma \sqrt{h} d^3x \, \epsilon^{abc}
\bigl( \mathcal{A}_a\partial_b \mathcal{A}_c^{(0)} - \mathcal{A}_a^{(0)}\partial_b \mathcal{A}_c \bigr), \label{LL}
\end{equation}
where $\mathcal{A}^{(0)}$ is a fiducial gauge field that has to be chosen in such a way that $\mathcal{A}_\alpha^{(0)}$ has the same Pontryagin number (\ref{Pont1}) as $\mathcal{A}_\alpha$ and does not depend on $x^3$ while $\mathcal{A}_3^{(0)}=a^{(0)}$ has to be a constant. One takes $a^{(0)}=\tfrac 12$ and $a^{(0)}=0$ for periodic and anti-periodic conditions on $S^1$, respectively. The formula (\ref{etaL}) then follows by inspection.

Suppose that $\mathcal{A}^+$ and $\mathcal{A}^-$ are of the type considered above. Since $\mathcal{A}^+-\mathcal{A}^-$ is a gauge vector, $\mathcal{A}_\alpha^\pm$ should have identical Pontryagin numbers on $\tilde\Sigma$. $\mathcal{A}_3^\pm =a^\pm$ are any constants. Then, by using either (\ref{etaper}) - (\ref{teta}) or (\ref{etaL}) and (\ref{LL}) one can easily check that the APS formula (\ref{APSwall}) holds. 

\section{Discussion and conclusions}\label{sec:dis}
In this paper, we have demonstrated that a suitably modified APS index theorem (\ref{APSwall}) holds for domain wall geometries in four dimensions that are characterized by gauge fields have a discontinuity at a submanifold $\Sigma$ of codimension one. We used the heat kernel techniques to compute the index and the $\eta$ functions. The general statement (\ref{APSwall}) was illustrated by two examples, a topologically trivial bundle and a topologically non-trivial abelian gauge field. We also discussed the variational properties of the $\eta$ functions and of the Chern-Simons action.

Whenever the Chern-Simons action is involved, gauge invariance is an issue to discuss. All spectral functions, as the index (\ref{Ind1}) and the $\eta$ function (\ref{etas}), are gauge invariant since they are constructed from gauge invariant spectra. For example, one may see gauge invariance of the index from Eqs.\ (\ref{a4}) that are written in terms of gauge covariant quantities. The function $\tilde\eta(\DD)$ is not invariant under large gauge transformations in general. In the example considered in Sec.\ \ref{sec:ex}, large gauge transformations with a winding number $p$ over $S^1$ change $a\to a+p$ and thus change $\tilde\eta(\DD)$. However, in the combination $\tilde\eta(\DD^+)+\tilde\eta(\DD^-)$ the gauge variations cancel, as it has to be due to the gauge invariance of the index. 

Let us discuss possible extensions of the results obtained. First of all, one has to remove the restrictions of vanishing Riemann curvature on $\MM$ and extrinsic curvature of $\Sigma$. Besides, it is interesting to see  whether one can allow for brane-world geometries (when the metric is continuous but not smooth across $\Sigma$). The heat kernel coefficients that are necessary for these cases can be found in \cite{Gilkey:2001mj}, but combinatorial complexity of the formulas grows significantly. Another extension may be to include more fields in the Dirac operator. Since $\slashed{D}$ has to anti-commute with $\gamma^5$, besides a gauge field and a metric with corresponding connection, one may only add an axial vector field, $\nabla_\mu \to \hat \nabla_\mu = \nabla_\mu +\ii \gamma^5V_\mu$. However, since $[\hat\nabla_\mu, \gamma^\nu ]\neq 0$ the original APS construction does not go through. Thus, an additional axial vector field probably does not lead to interesting results. Finally, one may consider more general matching conditions $\psi^-=T\psi^+$ with some transfer matrix $T$ instead of (\ref{match1}). At any rate, the most important task seems to be to understand the APS theorem for domain walls (\ref{APSwall}) without calculating the right and left hand sides separately.

\acknowledgments
I am grateful to Edward Witten for correspondence and to Maxim Kurkov for collaboration on related subjects.
This work was supported by the grants 2016/03319-6 and 2017/50294-1 of the S\~ao Paulo Research Foundation (FAPESP),  by the grant 303807/2016-4 of CNPq, by the RFBR project 18-02-00149-a and by the Tomsk State University Competitiveness Improvement Program.

\bibliographystyle{JHEP}
\bibliography{parity}

\end{document}